\documentclass[author-year,12pt]{article}
\usepackage{epsfig}
\usepackage{amssymb,mathrsfs,mathtools,enumerate}
\usepackage[normalem]{ulem}
\usepackage{amsmath}
\usepackage{bm}
\usepackage{xcolor}
\usepackage{lineno}
\usepackage{color}
\usepackage{soul}
\usepackage{graphicx}
\usepackage{epstopdf} 
\usepackage{wrapfig}
\usepackage{cite}

\newcommand{\diff}[2]{\dfrac{\mathrm{d}{#1}}{\mathrm{d}{#2}}}
\newcommand{\pdiff}[2]{\dfrac{\partial{#1}}{\partial{#2}}}
\makeatletter
\renewcommand\@biblabel[1]{#1.}
\makeatother 
\title{Reproduction Number And Asymptotic Stability For The Dynamics of a Honey Bee Colony with Continuous Age Structure}
\author{
	{\bf  M.I. Betti$^1$, L.M. Wahl$^{1,\dag}$, M. Zamir$^{1,2}$}
	\\$^1$Department of Applied Mathematics
	\\$^2$Department of Medical Biophysics 
	\\Western University 
	\\London, Ontario, N6A 5B7
	\\CANADA
	\\$^\dag$corresponding author: lwahl@uwo.ca
}
\begin{document}

\maketitle

\begin{abstract}
A system of partial differential equations is derived as a model for the dynamics of a honey bee colony with a continuous age distribution, and the system is then extended to include the effects of infectious disease.  In the disease-free case we analytically derive the equilibrium age distribution within the colony and propose a novel approach for determining the global asymptotic stability of a reduced model. Finally, we present a method for determining the basic reproduction number $R_0$ of the infection and the method can be applied to other age-structured disease models with interacting infected classes. The results of asymptotic stability indicate that a honey bee colony suffering losses caused by a hazard will recover naturally so long as the hazard is removed before the colony collapses. Our expression for $R_0$ has potential uses in the tracking and control of an infectious disease within a bee colony.
\end{abstract}

%\pagestyle{myheadings}
%\thispagestyle{plain}
%\markboth{M.I. Betti}{Asymptotic Stability and $R_0$ for Age-structured PDE Model}
 
 \section{Introduction}
 
 Honey bee populations continue to decline on a global scale \cite{Becher}, and as research efforts to identify the underlying cause or causes continue \cite{Dennis, Watanabe,Ho}, there is as yet no clear resolution of the problem. While the consequences of the decline are usually discussed in the context of agriculture and economics \cite{Calderone,Southwick,Neumann}, the key question clearly hinges on the stability of a honey bee colony as a population dynamical system. Mathematical models thus provide critically important tools for studying honey bee populations as they can both simulate many different environments as well as suggest potential sensitivities a colony may have to environmental hazards such as pesticides and climate change, or microbial hazards such as parasites and disease.

 Mathematical models have been used in the recent years to shed light on the effects of pesticides on the lifespan of foraging bees \cite{Khoury1,Khoury2}.  Other models have focused on the changing dynamics of a colony under changing environmental conditions \cite{Tereshko}, or the interactions between colonies \cite{Schmickl}. Recent work has also investigated  the effects of infection on honey bee colony dynamics, including Varroa mites \cite{Eberl1, Eberl,Eberl2} and nosema \cite{Betti1}. An approach from systems biology has been used to explore the multifactorial causes of colony failure \cite{Becher1}.
 
One of the complicating factors in the dynamics of a honey bee colony is the age distribution (structure) within the colony since age groups may differ in their expected lifespans, foraging behaviour or susceptibility to a given hazard. While age-independent models have provided a number of key insights into the properties of colony dynamics \cite{Betti1,Eberl,Khoury1}, the ultimate conditions for the survival or collapse of a honey bee colony must take into account this added dimension of the problem.

From a mathematical standpoint, incorporating the age structure of a honey bee colony into the equations governing the population dynamics leads to a set of partial differential equations instead of the ordinary differential equations obtained when age is not considered. The problem of finding stability and equilibrium conditions for the dynamics of the colony becomes correspondingly more complex. When considering disease dynamics, the problem of finding the basic reproduction number ($R_0$) at which the fate of the colony is at a bifurcation becomes particularly difficult.

We propose a method of resolving these mathematical difficulties by transforming the problem from the real space (age, time) into the Laplace space such that the governing equations become ordinary differential equations, and from the solution of these equations it is then possible to deduce the asymptotic behavior of the dynamical system in the real space. In particular, in the absence of disease we show that the system has an asymptotically stable equilibrium, which implies that the bee population will rebound from losses when hazards are removed, provided they are removed before the complete collapse of the colony. In the presence of disease we use a linearized form of the system to obtain a closed form expression for the basic reproduction number $R_0$. The expression thus provides a measurable threshold for whether the disease will decay or persist.

Determining $R_0$ for age-structured models has been previously studied for general disease models with one infected class \cite{Thieme, Allen, Hyman}. Our model has the added complexity of having two interacting infected classes, namely infected hive bees ($H_I$) and infected foragers ($F_I$), that cannot be transformed into a single infected class for analysis. Moreover, work on the asymptotic stability of disease models with age structure is sparse. Several studies use Lyapunov functions and semi-group analysis to prove persistence of solutions \cite{Magal, Hale}, while perturbation analysis has also proved useful \cite{Inaba}. Other work has made use of properties of the particular model at hand to show stability \cite{Castillo}.

In the present study, we take advantage of detailed experimental work elucidating the distinct roles that honey bees of different ages play within a hive, and use this information to develop a model of honey bee demographics with continuous age structure (Section 2).  This model is then extended to include the dynamics of an infectious disease within the colony. In particular, we find a stability threshold criterion which corresponds to the basic reproduction number R0. In Section 3.1, we use a steady-state approximation to derive the equilibrium age distribution of the disease-free hive.  In 3.2, we develop a novel Laplace transform approach to prove global stability of this disease-free distribution in a reduced model, implying local stability in the full model.  In 3.3, we again develop a novel approach to find a threshold criterion at which this equilibrium loses stability, and an infection will invade; this threshold corresponds to the basic reproductive number, $R_0$.  In Section 4,
we discuss the implications of these results and suggest possible applications of the new approaches we have developed to other problems in population dynamics.
 
 \section{Model}
  The proposed model combines the normal demographics of a honey bee colony with a disease that at first infects foragers and then spreads to the rest of the colony.  As in previous studies, to keep the problem tractable, we neglect the effects of the brood, guarding bees, as well as bees that work to repair the hive. Male honey bees, known as drones, do not contribute to the maintenance of the hive, so this population is also neglected. We thus focus on the hive bees, $H$, which are responsible for maintenance of the brood and the foragers, $F$, which are responsible for bringing food, $f$, into the hive. Generally, bees emerge from the brood as hive bees and are later recruited to foraging duties. These two classes are further divided, in the presence of disease, into susceptible populations, $H_S$ and $F_S$, and infected populations, $H_I$ and $F_I$. The following sections present the governing equations for each of these sub-populations. The total size of the colony to be modeled in age, $a$ (days), and time, $t$ (days), is thus,
  \begin{equation}
 \label{N}
 N(t)=\int(H_S(a,t)+H_I(a,t)+F_S(a,t)+F_I(a,t))\mathrm{d}a.
\end{equation} 
  
%%%%%%%%%%%%%%%%%%%%%%%%%%%%%%%%%%%%%%%%%%%%%%%%%%%%%%%%%%%
 \vglue 0.5cm
\textit{\textbf{Susceptible Hive Bees:}}  For the susceptible hive bee population $H_S(a,t)$, incorporating ag using the standard approach of McKendrick \cite{Mcken}, into an earlier formulation of the problem \cite{Betti1},  the governing equation becomes
 \begin{equation}
  \label{Hs}
  \dfrac{\partial H_S(a,t)}{\partial t}+\dfrac{\partial H_S(a,t)}{\partial a}=-u(a)H_S(a,t)-\beta I(t) H_S(a,t).
 \end{equation}
Here, the first term on the right-hand side describes the recruitment of hive bees to foraging duties, where $u(a)$ denotes the age-dependent rate of recruitment. Research has shown that juvenile hormone III regulates the age at which honey bees begin foraging \cite{Robinson}, and there is a minimum age, $a_R$, at which hive bees are normally recruited to foraging duties \cite{Fahrbach}. If the foraging needs of the colony are not being met, however, hive bees will be recruited to foraging duties at a younger age \cite{Huang}. Conversely, if the foraging bee population exceeds the needs of the colony, foragers produce a pheromone, ethyl oleate, to reduce recruitment \cite{Leoncini}. We incorporate these regulatory mechanisms by taking
 \begin{equation}
  \label{R}
  u(a)=\alpha\left(\dfrac{a}{a+k}\right)^2\left(1-\dfrac{\sigma}{N}{\int (F_S+F_I)\mathrm{d}a}\right)\mathrm{H_v}(a-a_R)
 \end{equation}
where $\alpha$ is a free parameter representing the base rate of recruitment, $\mathrm{H_v}(a-a_R)$ is the Heaviside function, and $1/\sigma$ is the maximum allowable fraction of foragers in the colony size. Thus recruitment begins at age $a_R$ and increases sigmoidally with age thereafter where $k$ is the age at which recruitment is at half its maximal rate.
The second term on the right-hand side of equation \eqref{Hs} governs the disease dynamics within the colony. 

We assume infection is transmitted via mass action at a constant rate $\beta$, and infection can be transferred from hive bees to foragers or vice versa. Such a mechanism can approximate transmission of a disease such as nosema \cite{Stevanovic}, which infects food stores and bees via the microsporidian \emph{Nosema ceranae} \cite{Botias, Smith}. Our mass action mechanism assumes that the amount of infected food is proportional to the number of infected bees who are handling said food. 

The total infected population of the hive, to be denoted by $I$, is then given by
\begin{equation}
 \label{Nfancy}
 I(t)=\int (H_I(a,t)+F_I(a,t))\mathrm{d}a.
\end{equation}

It is assumed that the hive provides sufficient safety for bees that remain within it \cite{Khoury1,Seeley2} such that the natural death rate of healthy hive bees is negligible compared to the rate of recruitment.
%%%%%%%%%%%%%%%%%%%%%%%%%%%%%%%%%%%%%%%%%%%%%%%%%%%%%%%%%%%%
\vglue 0.5cm
\textit{\textbf{Infected Hive Bees:}} Infected hive bees are at risk of dying due to disease at an age-dependent rate $d(a)$. The equation governing their dynamics is then
\begin{equation}
 \label{Hi}
 \dfrac{\partial H_I(a,t)}{\partial t}+\dfrac{\partial H_I(a,t)}{\partial a}=\beta I(t) H_S(a,t)-[u(a)+d(a)]H_I(a,t).
\end{equation}
%%%%%%%%%%%%%%%%%%%%%%%%%%%%%%%%%%%%%%%%%%%%%%%%%%%%%%%%%%%%
\vglue 0.5cm
\textit{\textbf{Susceptible Foragers:}} Susceptible foragers are recruited from susceptible hive bees and are subject to age-dependent natural death at rate $\mu(a)$. The equation governing their dynamics is thus given by
\begin{equation}
 \label{Fs}
 \dfrac{\partial F_S(a,t)}{\partial t}+\dfrac{\partial F_S(a,t)}{\partial a}=u(a)H_S(a,t)-[\mu(a)+\beta I(t)]F_S(a,t).
\end{equation}
%%%%%%%%%%%%%%%%%%%%%%%%%%%%%%%%%%%%%%%%%%%%%%%%%%%%%%%%%%%%
\vglue 0.5cm
\textit{\textbf{Infected Foragers:}} Infected foragers can be either (i) infected hive bees that have been recruited to foraging duties or (ii) susceptible foragers that have become infected by either infected foragers or infected hive bees. They are subject to a disease-related death rate, $d(a)$, and their dynamics are governed by
\begin{equation}
 \label{Fi}
 \dfrac{\partial F_I(a,t)}{\partial t}+\dfrac{\partial F_I(a,t)}{\partial a}=u(a)H_I(a,t)+\beta I(t)F_S(a,t)-[\mu(a)+d(a)]F_I(a,t).
\end{equation}
%%%%%%%%%%%%%%%%%%%%%%%%%%%%%%%%%%%%%%%%%%%%%%%%%%%%%%%%%%%%
\vglue 0.5cm
\textit{\textbf{Food stores:}} Food, $f$, is brought into the hive by both susceptible and infected foragers. For simplicity, we assume that all foragers bring in food at the same rate $c$ (g/day), although it is likely that infected foragers would be less efficient at the task. Food is consumed by foragers and hive bees at, again for simplicity, the same rate, $\gamma$. The amount of food available at time $t$ is therefore given by
\begin{equation}
 \label{f}
 \dfrac{\mathrm{d}f}{\mathrm{d}t}=c\int \left(F_S+F_I\right)\mathrm{d}a-\gamma N
\end{equation}
%%%%%%%%%%%%%%%%%%%%%%%%%%%%%%%%%%%%%%%%%%%%%%%%%%%%%%%%%%%%
\vglue 0.5cm
\textit{\textbf{Boundary Conditions:}} The system of equations \eqref{Hs},\eqref{Hi},\eqref{Fs},\eqref{Fi} is subject to the following boundary conditions:
\begin{align}
\label{BC}
 \begin{dcases}
  H_S(0,t)=LV\\
  H_I(0,t)=F_S(0,t)=F_I(0,t)=0\\
  \lim_{a\to\infty}H_S(a,t)=\lim_{a\to\infty}H_I(a,t)=\lim_{a\to\infty}F_S(a,t)=\lim_{a\to\infty}F_I(a,t)=0
 \end{dcases}
\end{align}
The first condition represents the birth of new bees where $L$ is the daily egg laying rate by the queen and $V$ is a survivability function which determines how many of the brood survive to become viable hive bees. The brood needs both sufficient food and sufficient care from the hive bees in order to survive \cite{Jones}. Moreover, it has been shown that there is a range of ages within which hive bees will care for the brood. Field data suggest that hive bees take on nursing duties at a minimum age $a_{mn}$ and complete these duties at a maximum age $a_{mx}$. After this age, hive bees tend to either transition to foraging duties or take on  hive security or maintenance duties \cite{Winston}. On this basis, we define the survivability function $V$ as
\begin{align}
 \label{S}
 V=\left(\dfrac{f}{b+f}\right)\left(\dfrac{\int_{a_{mn}}^{a_{mx}}H_S(a,t)\mathrm{d}a}{w+\int_{a_{mn}}^{a_{mx}}H_S(a,t)\mathrm{d}a}\right),
\end{align}
where $b$ is the amount of food required for half the brood to survive to adulthood in the presence of sufficient care from hive bees, and $w$ is the number of nursing hive bees necessary to ensure survival of half the brood in the presence of sufficient food stores.

Equations \eqref{Hs},\eqref{Hi},\eqref{Fs},\eqref{Fi},\eqref{f} form a system of integro-partial differential equations to be solved simultaneously subject to the boundary conditions \eqref{BC}.

%%%%%%%%%%%%%%%%%%%%%%%%%%%%%%%%%%%%%%%%%%%%%%%%%%%%%
\section{Results}
\subsection{Disease-free Equilibria (DFE)}
The factor $\sigma$ in equation \eqref{R} reduces the rate of recruitment to  foraging when the proportion of foragers in the colony approaches its optimal level. If we set $\sigma=0$ in the recruitment function, $u$, we are left with a linear system of partial differential equations. This is a reasonable approximation for steady state analysis, as at equilibrium, since the total number of foragers is fixed, the term
\[\left(1-\dfrac{\sigma}{N}{\int (F_S+F_I)\mathrm{d}a}\right)\] 
is constant and can be absorbed into $\alpha$. Since $\alpha$ can be scaled out via a rescaling of time, we set $\alpha=1$ for the rest of the analysis without loss of generality. Moreover, the equation for $H_S$ (equation \eqref{Hs}) is now decoupled from the equation for $F_S$ (equation \eqref{Fs}). We use this linear approximation to find the disease-free equilibria, DFE, for this system (i.e. equilibria which correspond to $H_I=F_I=0$). 

Numerical experiments suggest that there is only one true equilibrium for the system (equations \eqref{Hs},\eqref{Hi},\eqref{Fs},\eqref{Fi},\eqref{f}), namely the trivial case
\begin{equation}
 \label{triv}
 H_S(a)=F_S(a)=f(t)=0
\end{equation}
because as long as there are any hive bees or foragers, food stores continue to grow. This case may be interpreted as either the nonexistence or the total extinction of the colony.

Instead, therefore, we seek \textit{quasi-steady states} in which 
\begin{equation}
 \label{sstates}
 H_I(a,t)=F_I(a,t)=0,\quad f(t)\neq 0
\end{equation}
which correspond to nontrivial biologically viable states and are found by setting $\,\pdiff{H_S}{t}=\pdiff{F_S}{t}=0$ in equations \eqref{Hs},\eqref{Fs} to get:
\begin{align}
\label{odeH}
 \diff{H_S^*}{a}&=-u(a)H_S^*(a)\\
 \label{odeF}
 \diff{F_S^*}{a}&=u(a)H_S^*(a)-\mu(a)F_S^*(a)
\end{align}
where $H_S^*$ and $F^*_S$ denote the quasi-steady state solutions.

\begin{figure}[h]
	\centering
	\includegraphics[scale=0.6]{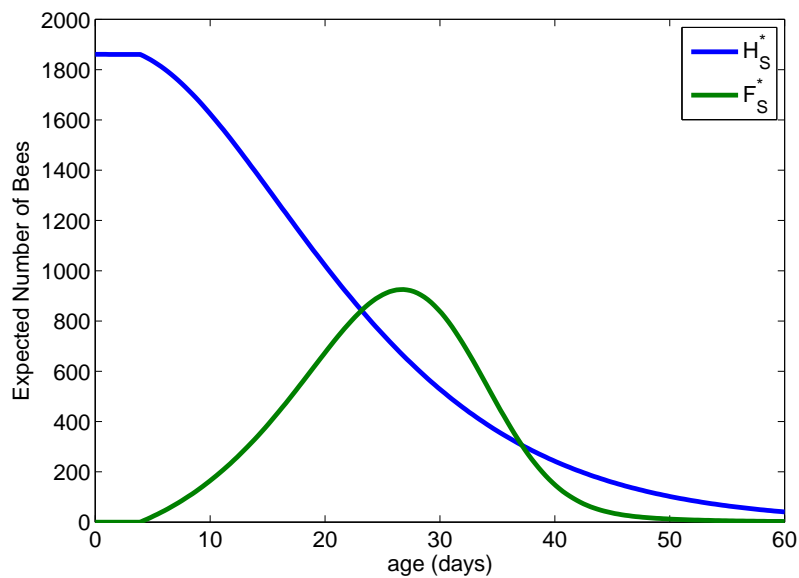}
	\caption{The disease-free equilibrium distributions $H_S^*$ and $F_S^*$} 
	\label{figEq}
\end{figure}

Equation \eqref{odeH} is decoupled from equation \eqref{odeF} and is linear. It can be integrated directly. Using the results, equation \eqref{odeF} can then be solved by variation of parameters, as it is a linear, non-homogeneous equation. The full solution to equations \eqref{odeH} and \eqref{odeF} is finally given by
\begin{equation}
\label{Heq}
 H_S^*(a)=\begin{dcases} H_{S0}, & \mbox{if } a<a_R \\ H_{{S0}}\dfrac{ \left( a+k \right) ^{2\,k}{{\rm exp}\left({-{\dfrac {{a}^{2}+ak-{k}^{2
}}{a+k}}}\right)}}{\left(  \left( a_{{R}}+k \right) ^{2\,k} \right) {{\rm exp}\left(-{\dfrac {{a_{{R}}}^{2}+a_{{R}}k-{k}^{2}}{a_{{R}}+k}}\right)
} }
, & \mbox{if } a\geq a_R \end{dcases}
\end{equation}
and 
\begin{equation}
 \label{Feq}
 F_S^*(a)=\begin{dcases} F_{{S0}}\,\,{{\rm exp}\left({\int _{0}^{a}\!-\mu \left( {\it a^*} \right) {d{\it 
a^*}}}\right)},&\mbox{if } a<a_R\\
\left(A(a)+F_{S0}\right){{\rm exp}\left({\int _{a_{{R}}}^{a}\!-\mu \left( {\it 
a^*} \right) {d{\it a^*}}}\right)},& \mbox{if } a\geq a_R
\end{dcases}
\end{equation}
where
\begin{equation}
\begin{dcases}
H_{S0}=H_S^*(0)\\
F_{S0}=F_S^*(0)
\end{dcases},
\end{equation}
\vglue 0.5cm
\begin{equation}
 \label{A}
 A(a)=  \int _{a_{{R}}}^{a}\! \dfrac{\left( {\it a^*}+k \right) ^
{-2+2\,k}{{\it a^*}}^{2}}{{k}^{2\,k}}H_{{S0}}E(a^*){d{\it a^*
}}
\end{equation}
and
\begin{align}
 E(a)={{\rm exp}\left({{\dfrac {-2\,{\it a}\,k-{{
\it a}}^{2}+(a+k)\int _{a_{{R}}}^{{\it a}}\!\mu \left( {\it a^*}
 \right) {d{\it a^*}}}{{\it a}+k}}}\right)}.
\end{align}

The conditions on `$a$' stem from the Heaviside term in the definition of $u(a)$ in equation \eqref{R}. The age distributions of susceptible hive and foraging bees in equations  \eqref{Heq} and \eqref{Feq} are shown in Figure \ref{figEq}.
%%%%%%%%%%%%%%%%%%%%%%%%%%%%%%%%%%%%%%%%%%%%%%%%%%%%%%%%%%%%%%%
\subsection{Stability of DFE}
If  the Laplace transform of $H(t)$ is denoted by $\displaystyle\mathscr{L}\{H(t)\}=\mathscr{H}(s)$ and $\displaystyle\lim_{t\to\infty}H(t)$ is finite, then by the Final Value Theorem \cite{McGraw}, 
 \begin{equation}
\lim_{t\to\infty}H(t)=\lim_{s\to0}s\mathscr{H}(s)
\end{equation}
 under two conditions \cite{Cannon}:
 \begin{enumerate}[i)]
  \item There is at most one simple pole at the origin in $s$-space.
  \item Any roots of the denominator of $\mathscr{H}(s)$ are negative.
 \end{enumerate}
In what follows we use this theorem to analyze the asymptotic behaviour of the PDE system given by equations \eqref{Hs}, \eqref{Hi}, \eqref{Fs}, \eqref{Fi} and \eqref{f}, and ultimately show that the disease-free equilibrium is stable.
 
We begin by taking the Laplace Transform in $t$ of equation \eqref{Hs}, ignoring the term $\beta IH_S$ since here we are interested in the disease-free case, which yields 
\begin{equation}
  \label{HLaplace}
  s\mathscr{H}(a,s)-H_S(a,0)+\dfrac{d \mathscr{H}(a,s)}{da}=-u(a)\mathscr{H}(a,s)
 \end{equation}
 where 
 \[\mathscr{H}(a,s)=\mathscr{L}\{H(a,t)\}.\]
The corresponding boundary condition is then given by taking the Laplace transform of condition \eqref{BC} but this is not easy because of the time dependence of that boundary condition. We simplify this problem by noting that food stores are unbounded and we are interested in the behaviour of the system as $t\to\infty$. On this basis we use the approximation
 \begin{equation}
  \lim_{t\to\infty}\dfrac{f}{b+f}\approx1.
 \end{equation}
As well, the second factor in the function $V$ (equation \eqref{S}) is always between zero and one and will be a constant at equilibrium which we take as $0\leq \kappa<1$. Using these approximations, the boundary condition for $\mathscr{H}$ becomes
 \begin{equation}
  \label{LBC}
  \mathscr{H}(0,s)=\dfrac{L\kappa}{s}.
 \end{equation}
 
Solving the ODE \eqref{HLaplace} with initial condition $H(a,0)=g(a)\geq0$ (bounded, analytic, and satisfying the two aforementioned conditions) yields,
 \begin{equation}
  \label{LaplaceSol}
  \mathscr{H}(a,s)= \begin{dcases} \left( \int _{0}^{a}\!g \left( {\it \tilde{a}} \right) {{\rm e}^{s{\it 
\tilde{a}}}}{d{\it \tilde{a}}}+{\dfrac {LW}{s}} \right) {{\rm e}^{-sa}} & a<a_R\\
                      \left( x(a)+C \right)  y(a) & a\geq a_R
                    \end{dcases}
 \end{equation}
 where 
 \begin{equation}
  y(a)=\left( a+k
 \right) ^{2\,k}{{\rm exp}\left({-{\dfrac {s{a}^{2}+sak+{a}^{2}+ak-{k}^{2}}{a+
k}}}\right)}
 \end{equation}
 \begin{equation}
  x(a)= \int _{a_{{R}}}^{a}\!g \left( {\it \tilde{a}} \right)  \left( {\it 
\tilde{a}}+k \right) ^{-2\,k}{{\rm exp}\left({{\dfrac {s{{\it \tilde{a}}}^{2}+s{\it \tilde{a}
}\,k+{{\it \tilde{a}}}^{2}+{\it \tilde{a}}\,k-{k}^{2}}{{\it \tilde{a}}+k}}}\right)}{d{\it 
\tilde{a}}}
 \end{equation}
 and
 \begin{equation}
  C=LW{s}^{-1} \left(  \left( a_{{R}}+k \right) ^{2\,k} \right) ^{-1
} \left( {{\rm exp}\left({-{\dfrac {s{a_{{R}}}^{2}+sa_{{R}}k+{a_{{R}}}^{2}+a_{
{R}}k-{k}^{2}}{a_{{R}}+k}}}\right)} \right) ^{-1}.
 \end{equation}
 The tilde sign signifies dummy variables.
 
We examine this result in two parts:

\underline{\textbf{(i) $\bm{a>a_R}$ :}} Here we observe that
 \begin{equation}
  \label{poscond}
  s \,y(a)x(a)\geq0
 \end{equation}
 and since $g(a)$ is bounded, and 
 \[
 0\leq\dfrac{\mathrm {exp}\left({-\dfrac{k^2}{a+k}}\right)}{(a+k)^{2k}}\leq1
  \]
 we find
 \begin{eqnarray}
 \label{upcond}
  s\,y(a)x(a)
&\leq&s\,y(a)\left(\max_{a\in[0,\infty)}g(a)\right)\int_{a_R}^a{\mathrm e}^{-a\left(\tilde{s}+1\right)}d\tilde{a}\\
\nonumber
  &=&s\,y(a)\left(\max_{a\in[0,\infty)}g(a)\right){\dfrac {{{\rm e}^{- \left( s+1 \right) {\it aR}}}-{{\rm e}^{- \left( s
+1 \right) a}}}{s+1}}.
 \end{eqnarray}
 We also note that 
 \begin{equation}
 \label{upper}
 \lim_{s\to0}s\,y(a)\left(\max_{a\in[0,\infty)}g(a)\right){\dfrac {{{\rm e}^{- \left( s+1 \right) {\it aR}}}-{{\rm e}^{- \left( s
+1 \right) a}}}{s+1}}=0.
\end{equation}

Using conditions \eqref{poscond}, \eqref{upcond}, \eqref{upper} and the Squeeze Theorem \cite{Stewart} we conclude that
\begin{equation}
 \label{flim}
 \lim_{s\to0}s\,y(a)x(a)=0
\end{equation}

As well, 
\begin{align}
\lim_{s\to0}s\, C&=\lim_{s\to0}\left\{sLW{s}^{-1} \left(  \left( a_{{R}}+k \right) ^{2\,k} \right) ^{-1
}\right. \\\nonumber &\phantom{ttttttttt}\left.\left( {{\rm exp}\left({-{\dfrac {s{a_{{R}}}^{2}+sa_{{R}}k+{a_{{R}}}^{2}+a_{
{R}}k-{k}^{2}}{a_{{R}}+k}}}\right)} \right) ^{-1}\right\}\\\nonumber
&=LW \left(  \left( a_{{R}}+k \right) ^{2\,k} \right) ^{-1} {\rm exp}\left(\dfrac {{a_{{R}}}^{2}+a_{{R}}k-{k}^{2}}{a_{{R}}+k} \right)\\
&=H_0.
\end{align}
 
Using the above, we find
\begin{eqnarray}
 \lim_{s\to0}s\mathscr{H}(a,s)&=&H_0\left( a+k
 \right) ^{2\,k}{{\rm exp}\left({-{\dfrac {s{a}^{2}+sak+{a}^{2}+ak-{k}^{2}}{a+
k}}}\right)}\\&=&H^*_S(a),
\end{eqnarray}
and applying the Final Value Theorem we find
\begin{equation}
 \lim_{t\to\infty}H_S(a,t)=H^*_S(a),\quad a\geq a_R
\end{equation}

\underline{\textbf{(ii) $\bm{a<a_R}$ :}} Here we note simply that
\begin{eqnarray}
 &&\lim_{s\to0}s\left( \int _{0}^{a}\!g \left( {\it \tilde{a}} \right) {{\rm e}^{s{\it 
\tilde{a}}}}{d{\it \tilde{a}}}+{\dfrac {LW}{s}} \right) {{\rm e}^{-sa}}\\&=&\left( \int _{0}^{a}\!\lim_{s\to0}sg \left( {\it \tilde{a}} \right) {{\rm e}^{s{\it 
\tilde{a}}}}{d{\it \tilde{a}}}+ \lim_{s\to0}{LW} \right) {{\rm e}^{-sa}}\\
&=&LW=H^*_S(a).
\end{eqnarray}

This completes the stability analysis of the DFE for $H_s$. Similar analysis applies to the DFE for $F_S$, though with considerably more tedious algebra, therefore we omit the details.

The above analysis determines the global stability of the linearized model in which we have sufficient food stores, brood care and time-independent recruitment (i.e. $\sigma=0$ in equation \eqref{R}). This linearized model is in fact an approximation to the full model governed by equations \eqref{Hs}, \eqref{Hi}, \eqref{Fs}, \eqref{Fi} and \eqref{f} near the quasi-steady state distribution given by equations \eqref{Heq} and \eqref{Feq}. Therefore, the global stability of the linearized model corresponds to  local stability of the full model.

The significance of these results is that starting from any initial age distribution $g(a)$, given sufficient food stores and brood care, a colony will rebound toward the distributions in equations \eqref{Heq} and \eqref{Feq}, shown numerically in Figure \ref{figEq}. While the analysis does not provide a time frame in which the rebound will occur, numerical experiments suggest that the rebound is relatively fast. More specifically, within months of an environmental hazard being removed the colony returns to its steady-state distribution.

%%%%%%%%%%%%%%%%%%%%%%%%%%%%%%%%%%%%%%%%%%%%%%%%%%%%%%%%%%%%%%%%%%%%%%%%%%%%%%%%%
\subsection{Basic Reproduction Number $\bm{R_0}$}
The basic reproduction number, $R_0$, for a system of partial differential equations has been explored previously in \cite{Thieme,Allen,Hyman}. The main difficulty in finding an expression for $R_0$ is that a system of partial differential equations has infinite dimensions and is therefore not amenable to standard methods such as the next generation matrix \cite{Diekmann,vdDriessche}. In the present case the situation is further compounded because there are two infective classes, namely $H_S$ and $F_S$, and they interact with each other. In what follows we propose techniques similar to those introduced by \cite{Allen} and \cite{Hyman} to determine the basic reproduction number in the face of these complications.

We begin by linearizing the system (equations \eqref{Hs},\eqref{Hi},\eqref{Fs},\eqref{Fi},\eqref{f}) about the DFE given by equations \eqref{Heq} and \eqref{Feq}. For the basic reproduction number, we are concerned with the growth of only the infected classes, governed by the following linearized equations:
\begin{align}
\label{HLin}
 \pdiff{H_I}{t}+\pdiff{H_I}{a}=&-\left(\left(\dfrac{a}{a+k}\right)^2\mathrm{H_v}(a-a_R)+d(a)\right)H_I\\\nonumber&+\beta H_S^*\int(H_I+F_I)\,\mathrm{d}a
 \end{align}
 \begin{align}
 \label{FLin}
 \pdiff{F_I}{t}+\pdiff{F_I}{a}=&\left(\dfrac{a}{a+k}\right)^2\mathrm{H_v}(a-a_R)H_I-(d(a)+\mu(a))F_I(a)\\\nonumber&+\beta F_S^*\int(H_I+F_I)\,\mathrm{d}a
\end{align}
where again we have assumed that $u(a)$ takes on the simplified form $\left(\dfrac{a}{a+k}\right)^2\mathrm{H_v}(a-a_R)$ at equilibrium. We then use the ansatz that at least for a short time after the infection begins, the two infected populations have fixed age distributions that grow or decay exponentially in time, that is
\begin{eqnarray}
 \label{ansatz1}
 H_I(a,t)&=&h_I(a)\mathrm{e}^{\rho t}\\
 \label{ansatz2}
 F_I(a,t)&=&f_I(a)\mathrm{e}^{\rho t}.
\end{eqnarray}
The set of all values of the exponent $\rho$ is thus viewed as the growth parameters of the linearized system, equations \eqref{HLin},\eqref{FLin}. For $\rho<0$ the solutions will decay to zero, and the DFE will be asymptotically stable. For $\rho>0$ the solutions will lead to an epidemic outbreak, and we now proceed to determine conditions under which this occurs.

Substituting \eqref{ansatz1} and \eqref{ansatz2} into \eqref{HLin} and \eqref{FLin} gives a system of ordinary differential equations:
\begin{eqnarray}
\label{hi}
 \diff{h_I}{a}&=&-\rho h_I-\left(\dfrac{a}{a+k}\right)^2u_{a_R}(a)h_I+b(a)W\\
 \label{fi}
 \diff{f_I}{a}&=&-\rho f_I+\left(\dfrac{a}{a+k}\right)^2u_{a_R}(a)h_I+B(a)W
\end{eqnarray}
where
\begin{equation}
 \label{W}
 W=\int \left(h_I+f_I\right)\mathrm{d}a,
\end{equation}
\begin{equation}
 \label{ba}
 b(a)=\beta H_S^*{(a)}
\end{equation}
and
\begin{equation}
 B(a)=\beta F_S^*{(a)}.
\end{equation}

The above system has the solution,
\begin{eqnarray}
 \label{hiSol}
 h_I(a)&=&W\mathrm{e}^{-\rho a}{{\rm e}^{-\Gamma_1(a)}}\int _{0}^{a}\!\mathrm{e}^{\rho \tilde{s}}b \left( {\tilde{s}}
 \right) {{\rm e}^{\Gamma_1(\tilde{s})}}{d{\tilde{s}}
}\\
 \label{fiSol}
 f_I(a)&=&W\mathrm{e}^{-\rho a}{{\rm e}^{-\Gamma_2(a)}}\int _{0}^{a}\!\mathrm{e}^{\rho \tilde{s}}{{\rm e}^{\Gamma_2(\tilde{s})}} \left( B
 \left( {\tilde{s}} \right) +u \left( {\tilde{s}} \right) {\it h_I^W}
 \left( {\tilde{s}} \right)  \right) {d{\tilde{s}}}
\end{eqnarray}
where
\begin{eqnarray}
\label{ua}
u(a)&=&\left(\dfrac{a}{a+k}\right)^2\mathrm{H_v}(a-a_R)\\
 \label{Gamma1}
 \Gamma_1(a)&=&\int _{0}^{a}\!u \left( {\tilde{a}} \right) +d \left( 
{\tilde{a}} \right) {d{\tilde{a}}}\\
\label{Gamma2}
\Gamma_2(a)&=&\int _{0}^{{a}}\!d \left( {\tilde{a}}
 \right) +\mu \left( {\tilde{a}} \right) {d{\tilde{a}}}\\
\label{hii}
h_I^W(a)&=&\mathrm{e}^{-\rho a}{{\rm e}^{-\Gamma_1(a)}}\int _{0}^{a}\!\mathrm{e}^{\rho \tilde{s}}b \left( {\tilde{s}}
 \right) {{\rm e}^{\Gamma_1(\tilde{s})}}{d{\tilde{s}}
}.
\end{eqnarray}

Substituting equations \eqref{hiSol} and \eqref{fiSol} into equation \eqref{W}, we find
\begin{equation}
\label{char}
 W=WR(\rho)
\end{equation}
where
\begin{eqnarray}
 \label{Rrho}
 R(\rho)=\int_0^{\infty}&\mathrm{e}^{-\rho a}&\left\{{{\rm e}^{-\Gamma_1(a)}}\int _{0}^{a}\!\mathrm{e}^{\rho s}b \left( {s}
 \right) {{\rm e}^{\Gamma_1(s)}}{d{s}
}\right.\\
\nonumber&&\left.+{{\rm e}^{-\Gamma_2(a)}}\int _{0}^{a}\!\mathrm{e}^{\rho s}{{\rm e}^{\Gamma_2(s)}} \left( B
 \left( {s} \right) +u \left( {s} \right) {\it h_I^W}
 \left( {s} \right)  \right) {d{s}}\right\}da.
\end{eqnarray}

To determine if a non-zero solution of equation \eqref{char} exists, we seek a $\rho_c$ such that
\begin{equation}
 R(\rho_c)=1.
\end{equation}

We use the Mean Value Theorem to simplify equation \eqref{Rrho} and rewrite equation \eqref{Rrho} as
\begin{eqnarray}
 \label{Rsimp}
 R(\rho)=&\int_0^{\infty}&\mathrm{e}^{\rho(c_1-a)}{{\rm e}^{-\Gamma_1(a)}}\int _{0}^{a} b \left( {s}\right) {{\rm e}^{\Gamma_1(s)}}{d{s}}\\
 \nonumber
 &+&\mathrm{e}^{\rho\left(c_2-a\right)}{{\rm e}^{-\Gamma_2(a)}}\int _{0}^{a}{{\rm e}^{\Gamma_2(s)}}  B
 \left( {s} \right)ds\\
 \nonumber
 &+&\mathrm{e}^{\rho\left(c_3-c_4-a\right)}\mathrm{e}^{-\Gamma_2(a)}\int_0^a u(s)\tilde{h}_I^W(s)\mathrm{e}^{\Gamma_2(s)}dsda
\end{eqnarray}
where
\begin{equation}
 \label{MVTcond}
 \left\{c_1,c_2,c_3,c_4\right\}\in[0,a]
\end{equation}
and, as a consequence,
\begin{equation}
 c_i-a\leq 0\quad\mathrm{for}\quad i=1,2,3,4.
\end{equation}
Taking the derivative of equation \eqref{Rsimp} with respect to $\rho$ yields
\begin{eqnarray}
\label{Rprime}
\,\,R^{\prime}(\rho)=-&\rho&\int_0^{\infty}\left\{|c_1-a|\mathrm{e}^{\rho(c_1-a)}{{\rm e}^{-\Gamma_1(a)}}\int _{0}^{a} b \left( {s}\right) {{\rm e}^{\Gamma_1(s)}}{d{s}}\right.\\
 \nonumber
 &+&|c_2-a|\mathrm{e}^{\rho\left(c_2-a\right)}{{\rm e}^{-\Gamma_2(a)}}\int _{0}^{a}{{\rm e}^{\Gamma_2(s)}}  B
 \left( {s} \right)ds\\
 \nonumber
 &+&\left.|c_3-c_4-a|\mathrm{e}^{\rho(c_3-c_4-a)}\mathrm{e}^{-\Gamma_2(a)}\int_0^a u(s)\tilde{h}_I^W(s)\mathrm{e}^{\Gamma_2(s)}ds\right\}da.
\end{eqnarray}

Recall that for a disease to persist, a value of $\rho_c>0$ is required such that $R(\rho_c)=1$. Since each integral in equation \eqref{Rprime} is positive (because each integrand is positive), then $R(\rho)$ is a non-increasing function of $\rho$. Therefore, if $R(0)<1$ then $R(\rho)\neq 1$ for any $\rho>0$. On the other hand, if $R(0)>1$ then by continuity of $R(\rho)$, there must exist a $\rho_c$ such that $R(\rho_c)=1$. This then implies there is a solution to equation \eqref{char} such that $W\neq0$ which then in turn implies that $h_I$ and $f_I$ are nonzero. Under these conditions an infection will persist.
\begin{quotation}
\noindent\textit{The above analysis provides the basis for taking $R(0)$ as our basic reproduction number, that is for setting $R(0)=R_0$. For $R_0<1$ the infection will decay, whereas for $R_0>1$ the infection will grow.} 
\end{quotation}
The basic reproduction number for this model is thus given by
\begin{align}
 R_0=\beta\int_0^{\infty}&\left\{\mathrm{e}^{-\Gamma_1(a)}\int_0^aH^*_S(s)\mathrm{e}^{\Gamma_1(s)}ds\right.\\\nonumber
 &\left.+\mathrm{e}^{-\Gamma_2(a)}\int_0^a\left(F^*_S(s)+u(s)\int_0^sH^*_S(\tilde{s})\mathrm{e}^{\Gamma_1(\tilde{s})}d\tilde{s}\right)\mathrm{e}^{\Gamma_2(s)}ds\right\}da,
\end{align}
which can also be written:
\begin{align}
 \label{R0}
 R_0=&\beta\int_0^{\infty}\left\{\int_0^aH^*_S(s)\mathrm{e}^{-\left(\Gamma_1(a)-\Gamma_1(s)\right)}ds\right\}da\\\nonumber
 &+\beta\int_0^{\infty}\left\{\int_0^aF^*_S(s)\mathrm{e}^{-\left(\Gamma_2(a)-\Gamma_2(s)\right)}ds\right\}da\\\nonumber
 &+\beta\int_0^{\infty}\left\{u(s)\mathrm{e}^{-\Gamma_2(a)}\int_0^sH^*_S(\tilde{s})\mathrm{e}^{-\left(\Gamma_2(s)-\Gamma_1(\tilde{s})\right)}d\tilde{s}ds\right\}da.
\end{align}
This value of $R_0$ defines a bifurcation point at which the disease-free equilibrium loses stability and gives rise to an epidemic. The appendix relates this expression for $R_0$ to the expression in the age independent case, $\hat{R}_0$. 

\begin{figure}[h]
	\centering
	\includegraphics[scale=0.6]{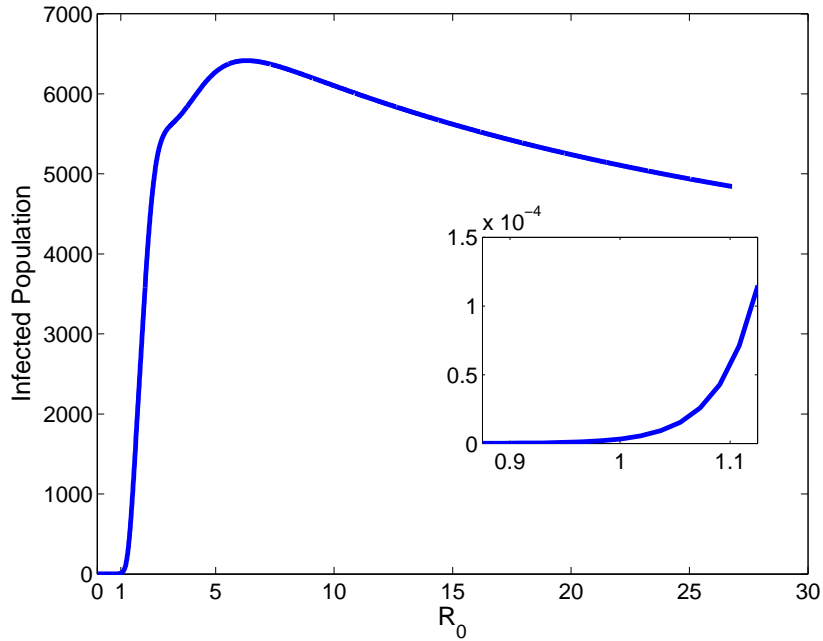}
	\caption{The total infected population in the hive after integrating equations \eqref{Hs},\eqref{Hi},\eqref{Fs} and \eqref{Fi} for 100 days vs. $R_0$ computed by numerical integration of equation \eqref{R0}. We see that the infection cannot infiltrate the colony for $R_0<1$.} 
	\label{fig1}
\end{figure}

Figure \ref{fig1} shows the total infected population, computed numerically and plotted against values of the basic reproduction number $R_0$. The figure confirms numerically the bifurcation value of $R_0$ at $1$ as predicted by the analysis. The reduction in $I$ seen as $R_0$ increases shows the slow collapse of the colony as the bee population becomes more afflicted by the disease.

\begin{figure}[h]
	\centering
	\includegraphics[scale=0.6]{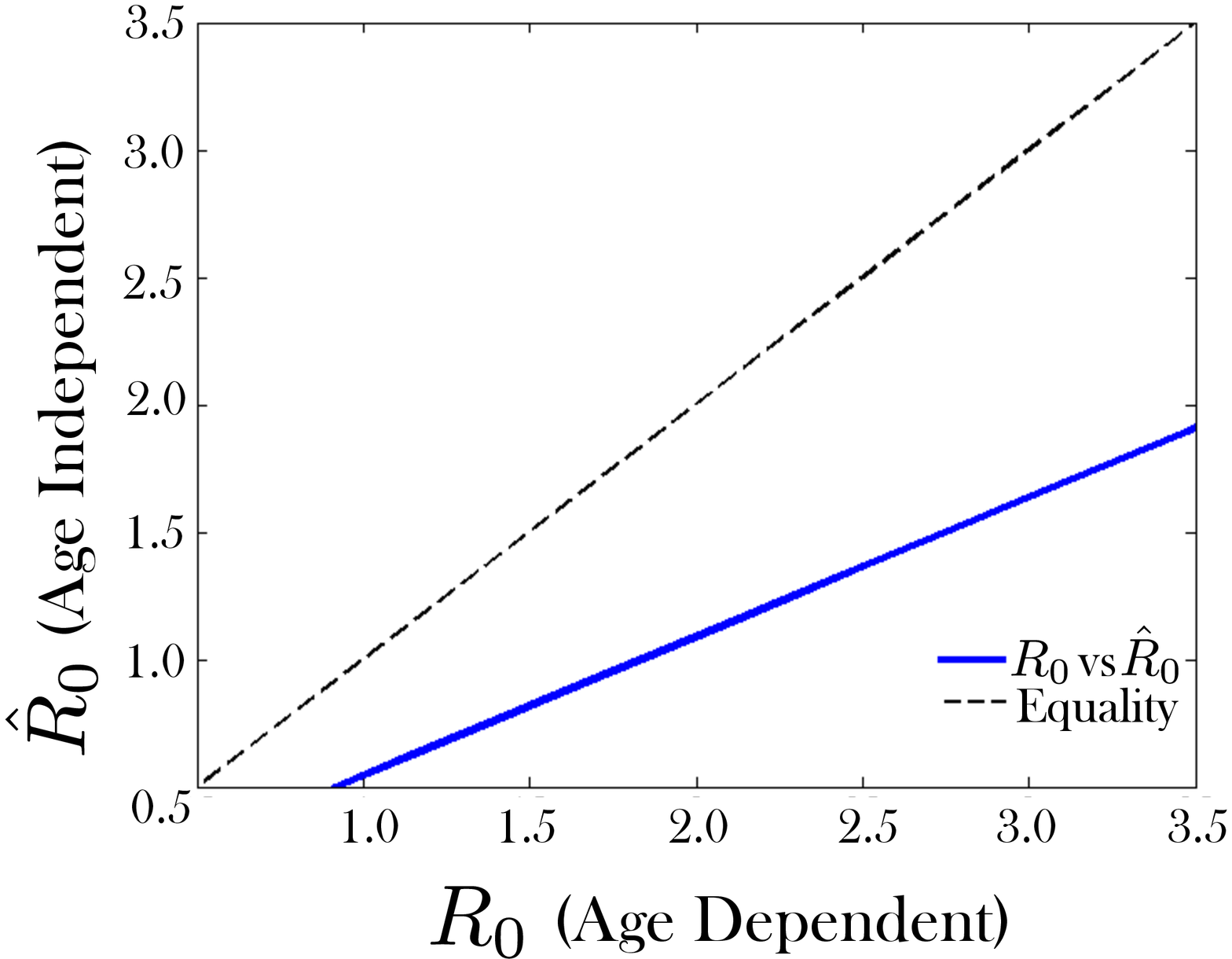}
	\caption{$R_0$ as calculated using equation \eqref{R0} plotted against the basic reproduction number in the age-independent case, $\hat{R}_0$ calculated using the next generation matrix. The dashed black line represents equality, while the data are shown in blue. The age-independent model consistently under estimates the value of $R_0$.}
	\label{fig2}
\end{figure}

Figure \ref{fig2} shows that the age-independent $R_0$ (where the parameter values are taken to be the average over all ages) significantly under-estimates the basic reproduction number. 

%%%%%%%%%%%%%%%%%%%%%%%%%%%%%%%%%%%%%%%%%%%%%%%%%%%%%%%%%%%%%%%%%%%%%%%%%%%%%%%%%
\section{Discussion}

An important prediction of this study is that if a hazard (environmental, parasitic, etc.) causes a bee colony to decline, the colony can recover if the hazard is removed in a timely fashion (given there are still sufficient food stores and brood care). Numerical experiments (data not shown) in fact suggest that the local stability result is quite robust and a colony is predicted to recover from significant losses. 

The results of this study also determine the necessary conditions under which the introduction of an infection will lead to an epidemic outbreak within a bee colony by determining the basic reproduction number, $R_0$. In theory, the formulation of $R_0$ can be verified using an infinite dimensional analog of the next generation matrix \cite{Thieme}. We confirm our derived expression of $R_0$ numerically and via reduction to an age-independent model, analyzed using the next generation matrix. Although many models have been proposed for disease dynamics within a honey bee colony \cite{Betti1, Eberl, Martin, Sumpter}, none have provided an expression for $R_0$. Here, we provide a closed-form expression for $R_0$ in both the age-independent case and for the continuous age-structured model. In theory, this should provide an upper and lower bound for $R_0$ for a model of $N$ age classes.

Moreover, $R_0$, although complicated in its formulation, can be used to determine whether a disease will be detrimental to a colony's health if the transmission rate and death rate associated with that disease are known. In practice, it is often easier to obtain an estimate of $R_0$ itself, in which case our formula can be used to determine the rate of transmission between bees or the distribution of the death rate. This knowledge may help in determining strategies for saving the colony.

As we saw, the basic reproduction number scales linearly with a constant transmission rate $\beta$, as expected. With similar analysis this result can be generalized to an age-dependent $\beta(a)$. Also, the rate of recruitment plays a role in the spread of infection to the extent that a change in rate of recruitment may have the potential to either increase or decrease $R_0$. In the case of the linearized, age-independent expression, $\hat{R_0}$, we find that an increase in recruitment lowers $R_0$ (see appendix). We also observe that by omitting age-structure from a model of honey bee colony dynamics, $R_0$ is under-estimated.

The method of using the Laplace transform to determine asymptotic stability can be used on more general time dependent systems of partial differential equations where the boundary conditions are constant. Also, the computation of $R_0$ is applicable to many biological models with interacting infected classes. For example, age-dependent models of many sexually transmitted diseases may have interacting infectious classes and could benefit from the approach we develop here.

In conclusion, the proposed model captures the critical role which age structure plays in the population dynamics of a honey bee colony specifically in the face of an infectious disease, thus aiding in the development of strategies for fighting the disease.

\section{Appendix}

\textit{\textbf{Test: Uniform age distribution. }}We test the validity of this bifurcation parameter by reducing equations \eqref{hi} and \eqref{fi} to a system in which all parameters are constant with respect to age. In doing so, we find from equation \eqref{R0} that 
\begin{equation}
\label{R0eqconst}
\hat{R_0}=\beta \dfrac{H_S^*}{u+d}+\beta\dfrac{uH_S^*}{\mu+d}+\beta\dfrac{F_S^*}{\mu+d}.
\end{equation}
This can be verified by using the next generation matrix on the infected classes of the following reduced model
\begin{eqnarray}
\label{consteq3}
\diff{H_S}{t}&=&-uH_S+\beta(H_I+F_I)H_S\\
\label{consteq4}
\diff{F_S}{t}&=&uH_S+\beta(H_I+F_I)F_S-\mu F_S\\
\label{consteq1}
\diff{H_I}{t}&=&-uH_I+\beta(H_I+F_I)H_S-dH_I\\
\label{consteq2}
\diff{F_I}{t}&=&uH_I+\beta(H_I+F_I)F_S-(d+\mu)F_I
\end{eqnarray}
which are a reduced form of equations \eqref{Hi} and \eqref{Fi}. The ratio of the disease-free equilibrium values of $F_S,H_S$ will always be such that 
\begin{align}
\label{ReducedRatio}
\dfrac{H_S^*}{F_S^*}=\dfrac{\mu}{u}.
\end{align}
This ratio is found by setting $H_I=F_I=\diff{H_S}{t}=\diff{F_S}{t}=0$ in equations \eqref{consteq3}, \eqref{consteq4}, \eqref{consteq1} and \eqref{consteq2}.

From these reduced equations we find the matrices,
\begin{align}
\label{F}
F=\left[\begin{array}{cc}
\beta H^*_S&\beta H^*_S\\
\beta F^*_S&\beta F^*_S
\end{array}\right]   
\end{align}
\begin{align}
\label{V}
V=\left[\begin{array}{cc}
d+u&0\\
-u&d+\mu
\end{array}\right]
\end{align}
which yield the next generation matrix
\begin{align}
\label{FV}
FV^{-1}=\left[\begin{array}{cc}
\dfrac{\beta H^*_S}{u+d}+\dfrac{\beta uH^*_S}{(u+d)(\mu+d)}&\dfrac{\beta H^*_S}{\mu+d}\\
\dfrac{\beta F^*_S}{u+d}+\dfrac{\beta uF^*_S}{(u+d)(\mu+d)}&\dfrac{\beta F^*_S}{\mu+d}
\end{array}\right]
\end{align}

Each term in this matrix has a biological interpretation which is the expected number of infections in each class ($H$ or $F$) caused by a single infected individual in each class. For example, the term
\begin{align}
\dfrac{\beta H^*_S}{u+d}
\end{align}
gives the expected number of susceptible hive bees that an infected hive bee will infect  while it is still a hive bee. The term
\begin{align}
\dfrac{\beta uH^*_S}{(u+d)(\mu+d)}
\end{align}
represents the probability that an infected hive bee will be recruited to foraging duties during its life time, multiplied by the expected number of susceptible hive bees that would then become infected. The expected number of susceptible hive bees infected by a single forager is given by
\begin{align}
\dfrac{\beta H_S^*}{\mu+d}
\end{align}

The interpretations for the second row of matrix \eqref{FV} are similar, but give the expected numbers of susceptible foragers that will become infected.

The basic reproduction number for this uniform age distribution model is then determined by the largest eigenvalue of the matrix $FV^{-1}$. Since we have the relation \eqref{ReducedRatio}, matrix \eqref{FV} is rank 1. Therefore, one of its eigenvalues is zero and the other is given by its trace. We can see that the trace of matrix \eqref{FV} gives the same expression for the basic reproduction number as \eqref{R0eqconst}.

The three terms that appear in \eqref{R0} are analogous to the three terms that appear in equation \eqref{R0eqconst}. This suggests that \eqref{R0} correctly determines not only the threshold for disease persistence but also correctly estimates the number of secondary infections subsequent to one primary infection \cite{Heffernan}.

%%%%%%%%%%%%%%%%%%%%%%%%%%%%%%%%%%%%%%%%%%%%%%%%%%%%%%%%%%%%%%%%
\bibliography{BeePopInfection}%%%%%%%%%%%%%%%

\begin{thebibliography}{10}

\bibitem{Allen}
{\sc L.~J. Allen, F.~Brauer, P.~Van~den Driessche, and J.~Wu}, {\em
  Mathematical epidemiology}, 2008.

\bibitem{ball1988prevalence}
{\sc B.~Ball and M.~Allen}, {\em The prevalence of pathogens in honey bee (apis
  mellifera) colonies infested with the parasitic mite varroa jacobsoni},
  Annals of applied biology, 113 (1988), pp.~237--244.

\bibitem{Becher1}
{\sc M.~A. Becher, V.~Grimm, P.~Thorbek, J.~Horn, P.~J. Kennedy, and J.~L.
  Osborne}, {\em Beehave: a systems model of honeybee colony dynamics and
  foraging to explore multifactorial causes of colony failure}, Journal of
  applied ecology, 51 (2014), pp.~470--482.

\bibitem{Becher}
{\sc M.~A. Becher, J.~L. Osborne, P.~Thorbek, P.~J. Kennedy, and V.~Grimm},
  {\em Review: Towards a systems approach for understanding honeybee decline: a
  stocktaking and synthesis of existing models}, Journal of Applied Ecology, 50
  (2013), pp.~868--880.

\bibitem{Betti1}
{\sc M.~I. Betti, L.~M. Wahl, and M.~Zamir}, {\em Effects of infection on honey
  bee population dynamics: A model}, PLoS one, 9 (2014), p.~e110237.

\bibitem{Betti3}
\leavevmode\vrule height 2pt depth -1.6pt width 23pt, {\em Age structure is
  critical to the population dynamics and survival of honey bee colonies}, Open
  Science,  (2016).
\newblock Accepted.

\bibitem{Botias}
{\sc C.~Botias, R.~Martin-Hernandez, L.~Barrios, A.~Meana, and M.~Higes}, {\em
  Nosema spp. infection and its negative effects on honey bees (\emph{{A}pis
  mellifera iberiensis}) at the colony level}, Veterinary research, 44 (2013),
  pp.~1--15.

\bibitem{Brauer}
{\sc F.~Brauer, C.~Castillo-Chavez, and C.~Castillo-Chavez}, {\em Mathematical
  models in population biology and epidemiology}, vol.~1, Springer, 2001.

\bibitem{Calderone}
{\sc N.~W. Calderone}, {\em Insect pollinated crops, insect pollinators and
  {US} agriculture: Trend analysis of aggregate data for the period
  1992-�2009}, PLoS ONE, 7 (2012), p.~e37235.

\bibitem{Cannon}
{\sc R.~H. Cannon}, {\em Dynamics of physical systems}, Courier Corporation,
  2003.

\bibitem{Castillo}
{\sc C.~Castillo-Chavez and Z.~Feng}, {\em Global stability of an age-structure
  model for {TB} and its applications to optimal vaccination strategies},
  Mathematical biosciences, 151 (1998), pp.~135--154.

\bibitem{Diekmann}
{\sc O.~Diekmann, J.~Heesterbeek, and J.~A. Metz}, {\em On the definition and
  the computation of the basic reproduction ratio ${R}_0$ in models for
  infectious diseases in heterogeneous populations}, Journal of mathematical
  biology, 28 (1990), pp.~365--382.

\bibitem{Dukas}
{\sc R.~Dukas}, {\em Mortality rates of honey bees in the wild}, Insectes
  Sociaux, 55 (2008), pp.~252--255.

\bibitem{Eberl1}
{\sc F.~M. R. K. P.~G. Eberl, Hermann~J.}, {\em Importance of brood maintenance
  terms in simple models of the honeybee - varroa destructor - acute bee
  paralysis virus complex.}, Electronic Journal of Differential Equations
  (EJDE) [electronic only], 2010 (2010), pp.~85--98.

\bibitem{Fahrbach}
{\sc S.~Fahrbach and G.~Robinson}, {\em Juvenile hormone, behavioral maturation
  and brain structure in the honey bee}, Developmental Neuroscience, 18 (1996),
  pp.~102--114.

\bibitem{Hale}
{\sc J.~K. Hale and P.~Waltman}, {\em Persistence in infinite-dimensional
  systems}, SIAM Journal on Mathematical Analysis, 20 (1989), pp.~388--395.

\bibitem{Heffernan}
{\sc J.~Heffernan, R.~Smith, and L.~Wahl}, {\em Perspectives on the basic
  reproductive ratio}, Journal of the Royal Society Interface, 2 (2005),
  pp.~281--293.

\bibitem{Ho}
{\sc M.-W. Ho and J.~Cummins}, {\em Mystery of disappearing honeybees}, Science
  in Society, 34 (2007), pp.~35--36.

\bibitem{Huang}
{\sc Z.-Y. Huang and G.~E. Robinson}, {\em Regulation of honey bee division of
  labor by colony age demography}, Behavioral Ecology and Sociobiology, 39
  (1996), pp.~147--158.

\bibitem{Hyman}
{\sc J.~M. Hyman and J.~Li}, {\em An intuitive formulation for the reproductive
  number for the spread of diseases in heterogeneous populations}, Mathematical
  biosciences, 167 (2000), pp.~65--86.

\bibitem{Inaba}
{\sc H.~Inaba}, {\em Threshold and stability results for an age-structured
  epidemic model}, Journal of mathematical biology, 28 (1990), pp.~411--434.

\bibitem{Jones}
{\sc J.~C. Jones, P.~Helliwell, M.~Beekman, R.~Maleszka, and B.~Oldroyd}, {\em
  The effects of rearing temperature on developmental stability and learning
  and memory in the honey bee, \emph{{A}pis mellifera}}, Journal of Comparative
  Physiology A, 191 (2005), pp.~1121--1129.

\bibitem{Khoury2}
{\sc D.~S. Khoury, A.~B. Barron, and M.~R. Myerscough}, {\em Modelling food and
  population dynamics in honey bee colonies}, PLoS ONE, 8 (2013), p.~e59084.

\bibitem{Khoury1}
{\sc D.~S. Khoury, M.~R. Myerscough, and A.~B. Barron}, {\em A quantitative
  model of honey bee colony population dynamics}, PLoS ONE, 6 (2011),
  p.~e18491.

\bibitem{Leoncini}
{\sc I.~Leoncini, Y.~Le~Conte, G.~Costagliola, E.~Plettner, A.~L. Toth, and
  M.~Wang}, {\em Regulation of behavioral maturation by a primer pheromone
  produced by adult worker honey bees}, Proceedings of the National Academy of
  Sciences of the United States of America, 101 (2004), pp.~17559--17564.

\bibitem{Magal}
{\sc P.~Magal, C.~McCluskey, and G.~Webb}, {\em Lyapunov functional and global
  asymptotic stability for an infection-age model}, Applicable Analysis, 89
  (2010), pp.~1109--1140.

\bibitem{mannshardt1978one}
{\sc R.~Mannshardt}, {\em One-step methods of any order for ordinary
  differential equations with discontinuous right-hand sides}, Numerische
  Mathematik, 31 (1978), pp.~131--152.

\bibitem{martin2000hygienic}
{\sc S.~Martin}, {\em Hygienic behaviour: an alternative view}, Bee
  Improvement, 7 (2000), pp.~6--7.

\bibitem{martin2001role}
{\sc S.~J. Martin}, {\em The role of varroa and viral pathogens in the collapse
  of honeybee colonies: a modelling approach}, Journal of Applied Ecology, 38
  (2001), pp.~1082--1093.

\bibitem{Mcken}
{\sc A.~McKendrick and M.~K. Pai}, {\em The rate of multiplication of
  microorganisms: a mathematical study}, in Proc. Roy. Soc. Edinburgh, vol.~31,
  1911, pp.~649--655.

\bibitem{Neumann}
{\sc P.~Neumann and N.~L. Carreck}, {\em Honey bee colony losses}, Journal of
  Apicultural Research, 49 (2010), pp.~1--6.

\bibitem{odoux2014ecobee}
{\sc J.-F. Odoux, P.~Aupinel, S.~Gateff, F.~Requier, M.~Henry, and
  V.~Bretagnolle}, {\em Ecobee: a tool for long-term honey bee colony
  monitoring at the landscape scale in west european intensive agroecosystems},
  Journal of Apicultural Research, 53 (2014), pp.~57--66.

\bibitem{McGraw}
{\sc S.~Parker}, {\em McGraw-Hill dictionary of scientific and technical
  terms}, McGraw-Hill, 2003.

\bibitem{perry2015rapid}
{\sc C.~J. Perry, E.~S{\o}vik, M.~R. Myerscough, and A.~B. Barron}, {\em Rapid
  behavioral maturation accelerates failure of stressed honey bee colonies},
  Proceedings of the National Academy of Sciences, 112 (2015), pp.~3427--3432.

\bibitem{petric2016mathematical}
{\sc A.~T. Petric}, {\em A Mathematical Model for a N. ceranae Infection in an
  A. mellifera Colony}, PhD thesis, 2016.

\bibitem{ratnieks1998queen}
{\sc F.~L. Ratnieks and L.~Keller}, {\em Queen control of egg fertilization in
  the honey bee}, Behavioral Ecology and Sociobiology, 44 (1998), pp.~57--61.

\bibitem{Eberl}
{\sc V.~Ratti, P.~G. Kevan, and H.~J. Eberl}, {\em A mathematical model for
  population dynamics in honeybee colonies infested with \emph{{V}arroa
  destructor} and the {A}cute {B}ee {P}aralysis {V}irus}, Canadian Applied
  Mathematics Quarterly: accepted, 21 (2013), pp.~63--93.

\bibitem{Eberl2}
\leavevmode\vrule height 2pt depth -1.6pt width 23pt, {\em A mathematical model
  of the honeybee-{V}arroa destructor-{A}cute {B}ee {P}aralysis {V}irus complex
  with seasonal effects}, Bulletin of Mathematical Biology,  (2015).

\bibitem{Robinson}
{\sc G.~E. Robinson, R.~E. Page, C.~Strambi, and A.~Strambi}, {\em Colony
  integration in honey bees: mechanisms of behavioral reversion}, Ethology, 90
  (1992), pp.~336--348.

\bibitem{Russell}
{\sc S.~Russell, A.~B. Barron, and D.~Harris}, {\em Dynamic modelling of honey
  bee ({A}pis mellifera) colony growth and failure}, Ecological Modelling, 265
  (2013), pp.~158 -- 169.

\bibitem{Sakagami}
{\sc S.~Sakagami and H.~Fukuda}, {\em Life tables for worker honeybees},
  Researches on Population Ecology, 10 (1968), pp.~127--139.

\bibitem{Schmickl}
{\sc T.~Schmickl and K.~Crailsheim}, {\em Hopomo: A model of honeybee
  intracolonial population dynamics and resource management}, Ecological
  modelling, 204 (2007), pp.~219--245.

\bibitem{seeley2009wisdom}
{\sc T.~D. Seeley}, {\em The wisdom of the hive: the social physiology of honey
  bee colonies}, Harvard University Press, 2009.

\bibitem{Seeley2}
{\sc T.~D. Seeley}, {\em Honeybee Democracy}, Princeton University Press, 2010.

\bibitem{Smith}
{\sc M.~L. Smith}, {\em The honey bee parasite \emph{{N}osema ceranae}:
  Transmissible via food exchange?}, PLoS ONE, 7 (2012), p.~e43319.

\bibitem{Southwick}
{\sc E.~E. Southwick and L.~Southwick~Jr}, {\em Estimating the economic value
  of honey bees (\emph{{H}ymenoptera: {A}pidae}) as agricultural pollinators in
  the {U}nited {S}tates}, Journal of Economic Entomology, 85 (1992),
  pp.~621--633.

\bibitem{Stevanovic}
{\sc J.~Stevanovic, P.~Simeunovic, B.~Gajic, N.~Lakic, D.~Radovic, I.~Fries,
  and Z.~Stanimirovic}, {\em Characteristics of \emph{{N}osema ceranae}
  infection in {S}erbian honey bee colonies}, Apidologie, 44 (2013),
  pp.~522--536.

\bibitem{Stewart}
{\sc J.~Stewart}, {\em Multivariable calculus}, Cengage Learning, 2011.

\bibitem{Sumpter}
{\sc D.~J.~T. Sumpter and S.~J. Martin}, {\em The dynamics of virus epidemics
  in {V}arroa-infested honey bee colonies}, Journal of Animal Ecology, 73
  (2004), pp.~51--63.

\bibitem{Tereshko}
{\sc V.~Tereshko and A.~Loengarov}, {\em Collective decision making in
  honey-bee foraging dynamics}, Computing and Information Systems, 9 (2005),
  p.~1.

\bibitem{Thieme}
{\sc H.~R. Thieme}, {\em Spectral bound and reproduction number for
  infinite-dimensional population structure and time heterogeneity}, SIAM
  Journal on Applied Mathematics, 70 (2009), pp.~188--211.

\bibitem{vdDriessche}
{\sc P.~Van~den Driessche and J.~Watmough}, {\em Reproduction numbers and
  sub-threshold endemic equilibria for compartmental models of disease
  transmission}, Mathematical biosciences, 180 (2002), pp.~29--48.

\bibitem{vdSteen}
{\sc J.~J. van~der Steen, B.~Cornelissen, J.~Donders, T.~Blacqui{\`e}re, and
  C.~van Dooremalen}, {\em How honey bees of successive age classes are
  distributed over a one storey, ten frames hive}, Journal of Apicultural
  Research, 51 (2012), pp.~174--178.

\bibitem{Dennis}
{\sc D.~vanEngelsdorp, J.~D. Evans, C.~Saegerman, C.~Mullin, E.~Haubruge, B.~K.
  Nguyen, M.~Frazier, J.~Frazier, D.~Cox-Foster, Y.~Chen, R.~Underwood, D.~R.
  Tarpy, and J.~S. Pettis}, {\em Colony collapse disorder: A descriptive
  study}, PLoS ONE, 4 (2009), p.~e6481.

\bibitem{wang2012basic}
{\sc W.~Wang and X.-Q. Zhao}, {\em Basic reproduction numbers for
  reaction-diffusion epidemic models}, SIAM Journal on Applied Dynamical
  Systems, 11 (2012), pp.~1652--1673.

\bibitem{Watanabe}
{\sc M.~E. Watanabe}, {\em Colony collapse disorder: Many suspects, no smoking
  gun}, BioScience, 58 (2008), pp.~384--388.

\bibitem{Winston}
{\sc M.~Winston}, {\em The biology of the honey bee}, Harvard University Press,
  1987.

\end{thebibliography}
\bibliographystyle{siam}
%%%%%%%%%%%%%%%%%%%%%%%%%%%%%%%%%%%%%%%%%%%%%%%%%%

%%%%%%%%%%%%%%%%%%%%%%%%%%%%%%%%%%%%%%%%%%%%%%%%%%%%

%%%%%%%%%%%%%%%%%%%%%%%%%%%%%%%%%%%%%%

\end{document}